\documentclass[aps,prl,reprint,groupedaddress,nofootinbib]{revtex4-1}

\usepackage{bbm}
\usepackage{amssymb}
\usepackage{graphicx}
\usepackage{amsmath,amsfonts,amsthm,bm} 
\usepackage{mathtools}
\usepackage{times}
\usepackage{bm}
\usepackage{natbib}
\usepackage{url}
\usepackage{xcolor}
\usepackage{multirow}
\usepackage{tabularx, ragged2e, booktabs}
\newcolumntype{Y}{>{\centering\arraybackslash}X}
\usepackage{adjustbox}
\usepackage{cprotect} 

\graphicspath{
{../../../MATLAB/Exercise/Freidlin_Wentzell/}
{../../../MATLAB/Useful/quasi_potential/}
}

\usepackage[update,prepend]{epstopdf} 

\begin{document}

\title{Transitions across Melancholia States in a Climate Model: Reconciling the Deterministic and Stochastic Points of View}

\author{\large Valerio Lucarini$^{1,2,3}$ and Tam\'as B\'odai$^{1,2}$\\
$^1$Centre for the Mathematics of Planet Earth, University of Reading, Reading, UK\\ 
$^2$Department of Mathematics and Statistics, University of Reading, Reading UK\\
$^3$CEN, University of Hamburg, Hamburg, Germany\\
}

\begin{abstract}

The Earth is well-known to be, in the current astronomical configuration, in a regime where two asymptotic states can be realised. The warm state we live in is in competition with the ice-covered snowball state. The bistability exists as a result of the positive ice-albedo feedback. In a previous investigation performed on a intermediate complexity climate model we have identified the {\color{black}unstable climate} states (\textit{Melancholia states}) separating the co-existing climates, and studied their dynamical and geometrical properties. The Melancholia states are ice-covered up to the mid-latitudes {\color{black}and attract} trajectories initialised  on the basins boundary. In this paper, we study {\color{black}how stochastically perturbing  the parameter controlling the intensity of the incoming solar radiation impacts} the stability of the climate. We detect transitions between the warm and the snowball state and analyse in detail the properties of the noise-induced escapes from the corresponding basins of attraction. We {\color{black}determine} the most probable paths for the transitions and find evidence that the Melancholia states act as gateways, similarly to saddle points in an energy landscape.


\end{abstract}
\maketitle
The Earth{\color{black}, for a vast range of parameters controlling its radiative budget, \textit{e.g.} the intensity of the solar irradiance and the concentration of greenhouse gases, including the present-day astronomical configuration and atmospheric composition}, supports two co-existing climates. One is the warm state we live in, and the other one is the \textit{snowball state}, featuring global glaciation and extremely low surface temperatures \cite{Hoffman2002,Pierrehumbert2011}. {\color{black} Indeed, events of onset and decay of snowball conditions have taken place in the Neoproterozoic  \cite{donnadieu2014}.}  The bistability of the climate system comes from the competition between the positive ice-albedo feedback ({\color{black}ice reflects efficiently the solar radiation}) and the negative Boltzmann feedback (a warmer surface emits more radiation), with the tippings point realised when the {\color{black}negative and positive feedbacks are equally strong, with ensuing loss of bistability}.  
With simple models one can identify, within the bistability region, unstable solutions{\color{black} - \textit{Melancholia states} -} sitting in-between the two stable climates. Melancholia (M) states are, {\color{black}far from the tipping points}, ice-covered up to the mid-latitudes. Small perturbations applied to trajectories initialised on the M states lead to the system falling  into either asymptotic state \cite{Budyko1969,Sellers1969,Ghil1976}. Improving our understanding of the related critical transitions is a key challenge for geoscience and has strong implications in terms of planetary habitability \cite{Pierrehumbert2011,LucAstr2013,Boschi2013,Lucarini2017}.  

The goal of this letter is to explore, using a simplified yet Earth-like climate model, the phase space   of the climate system {\color{black}by taking advantage of the rich dynamics resulting from adding stochastic perturbations}, and, in particular, by focusing on noise-induced transitions between the warm (W) and snowball (SB) attractors and linking this with the global stability properties  analysed in \cite{Lucarini2017} using tools and ideas of high-dimensional deterministic dynamical systems. {\color{black}The methodology  proposed here is of general relevance for studying multistable systems \cite{Feudel2018} and, specifically, for studying in a novel way the properties of the Earth tipping elements \cite{Lenton2008}.}
 
Multistable systems are extensively investigated both in natural and social sciences \cite{Feudel2018} and they can be introduced as follows. 
We consider a smooth autonomous continuous-time dynamical system acting on a smooth finite-dimensional compact manifold $\mathcal{M}$.  We define $\bm{x}(t,\bm{x}_0)$=$S^t(\bm{x}_0)$ as an orbit at time $t$, where $S^t$ is the evolution operator, and $\bm{x}_0$ the initial conditions at $t=0$.  We write the corresponding set of ordinary differential equations as ${\dot{\bm{x}}}=\bm{F}(\bm{x})$ 
where $\bm{F}(\bm{x})=d/d\tau S^\tau(\bm{x})|_{\tau=0}$ is a smooth vector field. {\color{black}The system is multistable if it possesses} more than one asymptotic states, defined by the attractors $\Omega_j$, $j=1,\ldots,J$. The  asymptotic state of the orbit is determined by its initial condition, and the phase space is partitioned between the basins of attraction $B_j$ of the  attractors $\Omega_j$ and the boundaries $\partial B_l$, $l=1,\ldots,L$ {\color{black}separating} such basins. If the dynamics is determined by the energy landscape $U(\bm{x})$, {\color{black}with} $\bm{F}(\bm{x})=-\bm{\nabla} U(\bm{x})$, the attractors {\color{black}are} the local minima $\bm{x}_j$, $j=1,\ldots,J$ of $U(\bm{x})$, the basin boundaries {\color{black}$\partial B_l$, $l=1,\ldots,L$} are the \textit{mountain crests}, which are smooth manifolds, {\color{black}each possessing} 
a minimum energy saddle - a \textit{mountain pass}.

{\color{black}More generally,}the basin boundaries can be strange geometrical objects with co-dimension smaller than one. {\color{black}Orbits initialized on the basin boundaries $\partial B_l$, $l=1,\ldots,L$ are attracted towards invariant saddles. 
Such saddles 
 $\Pi_l$, $l=1,\ldots,L$} can  feature chaotic dynamics  \cite{Grebogi1983,Robert2000,Ott2002,Vollmer2009}. 
 The definition of the basin boundaries and of the  {\color{black}saddles} is key for 
understanding the global stability properties of the system and its global bifurcations. 

\begin{figure}
\includegraphics[trim=1cm 0cm 0cm 1.22cm, clip=true, width=0.47\textwidth]{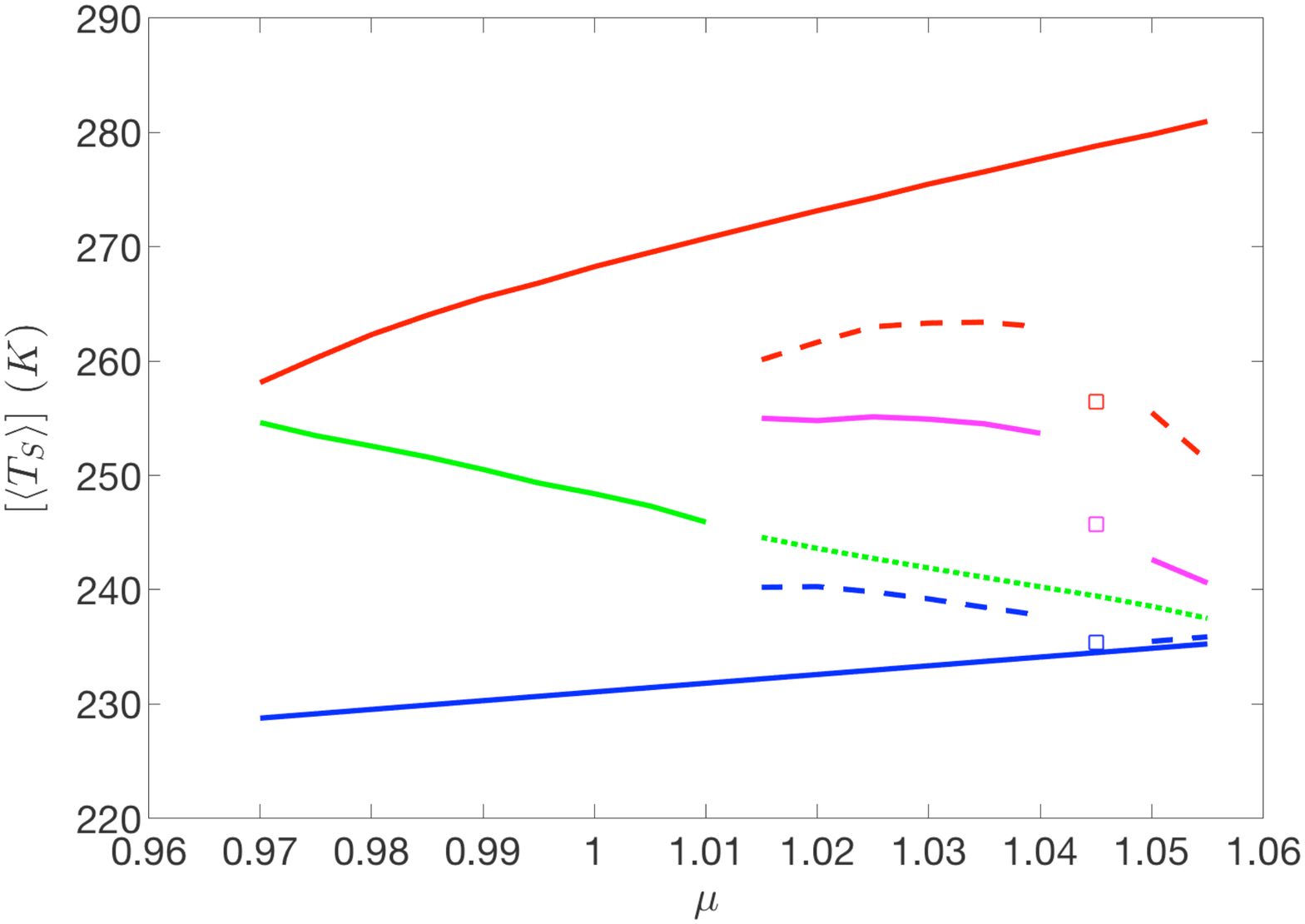}%
\caption{Bifurcation diagram for the   model studied in \cite{Lucarini2017} for the long term, globally averaged ocean temperature $[\langle T_S\rangle]$.  
 Bistability is found for a large range of values of the control parameter $\mu$. 
Of interest here: red {\color{black}line}: warm (W) states; blue {\color{black}line}: Snowball (SB) state; green {\color{black}line}: Melancholia (M) states, {\color{black}(constructed via the edge tracking algorithm}). The {\color{black}$W\rightarrow SB$ ($SB\rightarrow W$) tipping point is located at $\mu_{W\rightarrow SB}\sim 0.965$ ($\mu_{SB\rightarrow W}\sim 1.06$)}.}\label{fig1}
\end{figure}

In \cite{Bodai2014,Lucarini2017}, we adapted the {\color{black}edge tracking algorithm} presented in \cite{Skufca2006,Schneider2007} and {\color{black} constructed} in the bistable region the M states separating the two co-existing realisable W and SB climates. The critical transitions are {\color{black}  associated to boundary crises \cite{Ott2002} associated to collisions between the M state and one of the stable climates, and are flagged by a diverging linear response  \cite{Tantet2018}.} 
In \cite{Lucarini2017} we  constructed the M states for an intermediate complexity  climate model with O($10^5$) degrees of freedom.  We showed that the M state has, in a range of values of the control parameter $\mu$ ({\color{black} ratio between the considered solar irradiance and} the present-day {\color{black}value}), chaotic dynamics,  {\color{black} leading to weather variability and to} a limited horizon of predictability; see {\color{black} the caption of }Fig. \ref{fig1}. Since this instability is much faster than the climatic one due to the ice-albedo feedback, the basin boundary is a fractal set with near-zero codimension{\color{black}, {\color{black}in agreement with results obtained in low-dimensional cases \cite{Grebogi1983,LT:2011}}}. Near the basin boundary there is {\color{black}\textit{de facto}} no Lorenz' \cite{Lorenz1975} predictability of the second kind\footnote{In a a small window of values of $\mu$ we  discovered three stable states, with ensuing existence of multiple M states, with various possible topological configurations. This will not be discussed here. {\color{black}Yet, our approach can be adapted for dealing with systems with more than two stable states.}}.

 Here, building on \cite{Lucarini2017}, we study how a {\color{black}fluctuating} solar irradiance can trigger transitions between the W and SB states, {\color{black}and investigate the typical paths of such transitions}.  
The climate model  is constructed by coupling the primitive equations atmospheric model PUMA \cite{puma} with the Ghil-Sellers energy balance model \cite{Ghil1976} {\color{black}(see also \cite{Schneider1973,Dwyer1973})}, which describes succinctly {\color{black}oceanic heat transports}. The ocean model describes effectively the ice-albedo feedback, and defines the slow manifold of the system. The coupling is {\color{black}realised} by relaxing the atmospheric temperature to an adiabatic profile anchored to the ocean surface temperature, and by incorporating vertical heat fluxes. {\color{black}The ocean temperature  $T_S(t,\phi,\lambda)$, where 
$\phi$ is latitude and $\lambda$ is longitude, evolves as follows}:  
\begin{align}
C(\phi)\frac{\partial T_S}{\partial t}&= \mu(1+\sigma \frac{dW}{dt})I(\phi)\frac{S^*}{4}(1-\alpha(\phi,T_S))-O(T_S)\nonumber\\
&-D_\phi[T_S]+\chi[T_S,T_A],\label{1DEBM}
\end{align}
where $S^*$ is the present{\color{black}-day} solar irradiance (the factor 4 comes from the Earth-Sun geometry \cite{saltzman_dynamical}), 
and  the heat capacity $C$ and the geometrical factor $I$ depend explicitly on $\phi$. The albedo $\alpha$  depends  on $\phi$ and, critically, on $T_S$, with a rapid transition from {\color{black}high} albedo for low values of $T_S$ ($\alpha_{max}=0.6$) to {\color{black}low} albedo  for $T_S\gtrsim260$ $K$ ($\alpha_{min}=0.2$), which fuels the  ice-albedo feedback. Finally, $O$ is the outgoing radiation, {\color{black}}, and increases with $T_S$, ({\color{black}accounting for} the  Boltzmann feedback {\color{black}beside the greenhouse effect}), $D$ is a diffusion operator parametrizing the meridional heat transport, and $\chi$ describes the ocean-atmosphere heat exchange \cite{Lucarini2017}.

The stochastic perturbation {\color{black}modulates} the solar irradiance {\color{black}through} the factor  $(1+\sigma dW/dt)$, where $\sigma$ controls the noise intensity  noise, and $dW$ is the increment of a Wiener process. 
The {\color{black}noise is multiplicative because} $dW/dt$ is multiplied {\color{black} by} the factor $1-\alpha(\phi,T_S)$ in Eq. \ref{1DEBM}. Adding a Gaussian random variable of variance $\sigma_0$ at each time step $\Delta t$ ($1$ hour){\color{black}, when numerically integrating the model,  corresponds to having a relative fluctuation of the solar irradiance $\sigma_\tau=\sigma_0/\sqrt{N}$ on the time scale $\tau=N\times \Delta t$}. 



{\color{black}Noise-induced escape{\color{black}s} from  attractors} have been 
intensely studied \cite{Hanggi1986,Kautz1987,Grassberger1989}. {\color{black}We frame our problem by considering a stochastic differential equation in It\^o form written as
${d{\bm{x}}(t)}=\bm{F}(\bm{x(t)})dt+\sigma\bm{s}(\bm{x})d\bm{W}$,
where $\dot{\bm{x}}(t)=\bm{F}(\bm{x(t)})$ has multiple steady states, $d\bm{W}$ is the increment of an $M-$dimensional Wiener process, $\bm{s}(\bm{x})^T\bm{s}(\bm{x})$ is the noise covariance matrix with $\bm{s}(\bm{x}) \in \mathbb{R}^{N\times M}$, and $\sigma\geq 0 $. In the case of non-degenerate additive noise and for a class of multiplicative noise laws, the Freidlin-Wentzell~\cite{Freidlin1984} theory and extensions thereof \cite{Graham1991,Hamm1994,LT:2011} show {\color{black}that in the weak-noise limit $\sigma\rightarrow 0 $ the invariant measure  can be written}  as a large deviation law:
\begin{equation}\label{eq:stationary_distr}
  W_\sigma(\bm{x}) \sim Z(\bm{x})\exp\left(-\frac{2\Phi(\bm{x})}{\sigma^2}\right),
\end{equation}
where $Z(\bm{x})$ is the pre-exponential factor and $\Phi(\bm{x})$ is the pseudo-potential \footnote{$\Phi(\bm{x})=U(\bm{x})$ if $\bm{F}(\bm{x})=-\bm{\nabla}U(\bm{x})$ and $\bm{s}(\bm{x})^T\bm{s}(\bm{x})=\bm{1} \in  \mathbb{R}^{N\times N}$.}, which has local minima} at the deterministic attractors $\Omega_j$, $j=1,\ldots, J$. 
Both the M states and the attractors can be chaotic: if so, $\Phi$ has constant value over each M state and each attractor, respectively \cite{Graham1991,Hamm1994}. 
The probability that an orbit with initial condition in $B_j$ does not escape from it over a time {\color{black} $p$ decays as:
\begin{equation}\label{eq:tt_distr}
 P(p) =\frac{1}{\bar{\tau}_\sigma}\exp\left(-\frac{p}{\bar{\tau}_\sigma}\right), \quad \bar{\tau}_\sigma \propto \exp\left(\frac{2\Delta\Phi}{\sigma^2}\right)
\end{equation}
where $\bar{\tau}_\sigma$ is the expected }escape time and 
%
where $\Delta\Phi=\Phi(\Pi_l)-\Phi(\Omega_j)$ is the {\color{black}pseudo-}potential barrier height \cite{LT:2011}; {\color{black}In general, one may need to add a correcting prefactor in Eq. \ref{eq:tt_distr} \cite{LT:2011}.} 
In the weak-noise limit, the transition paths follow the instantons, which are minimizers of the Freidlin-Wentzell action \cite{Kautz1987,Grassberger1989,Kraut2002,Beri2005}. An instanton connects a point in $\Omega_j$ to a point in $\Pi_l$; if these sets are not fixed points, the instanton is not unique, as the {\color{black}pseudo-}potential is constant on $\Omega_j$ and $\Pi_l$. 


\begin{figure}
\includegraphics[trim=0cm 0.5cm 0cm 1cm, clip=true, width=0.48\textwidth]{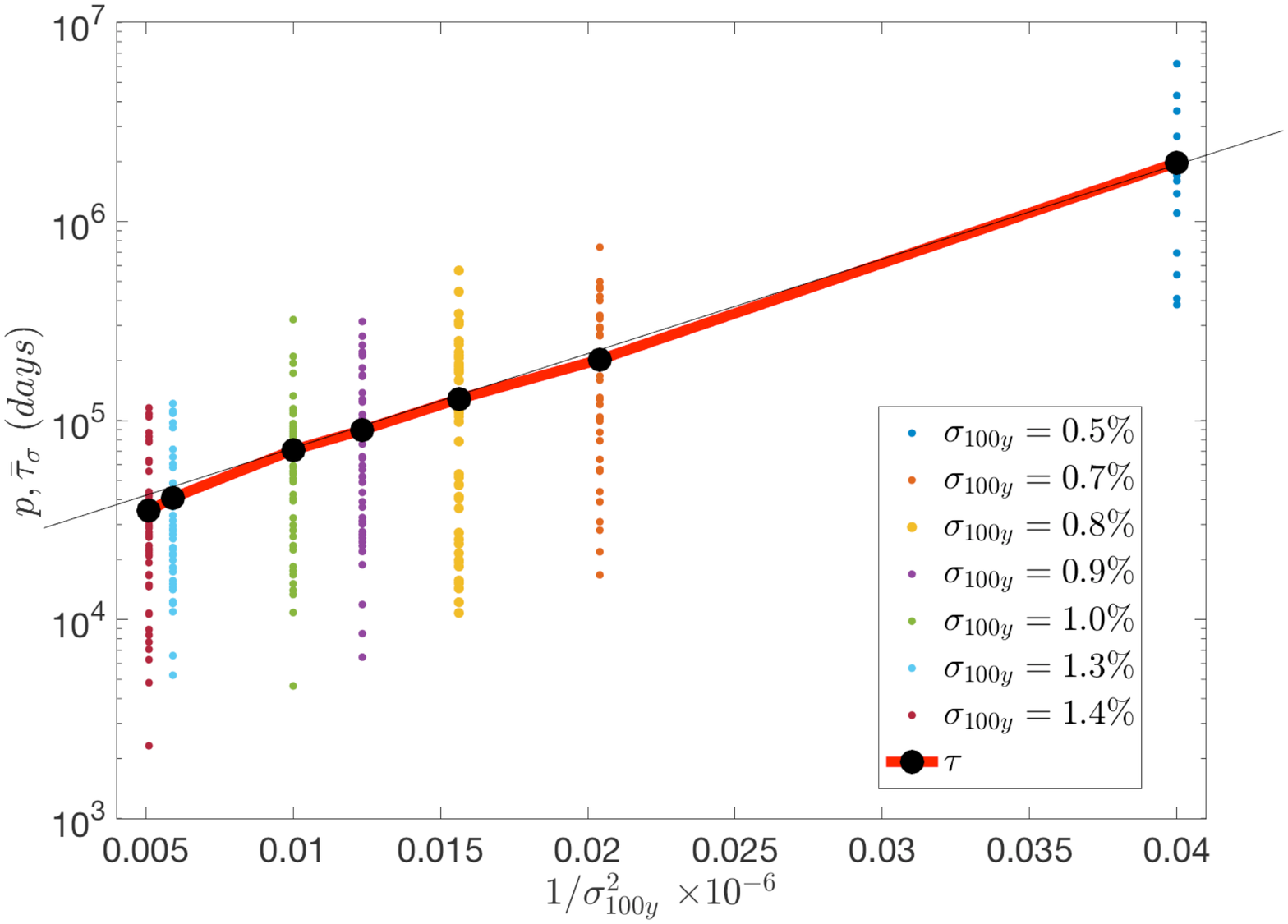}%
\caption{Escape times for the {\color{black}$W\rightarrow SB$} transitions for various noise strengths. Each dot corresponds to an observed escape time. The slope of the straight line fit is {\color{black}twice the quantity $\Delta \Phi$, see }Eq. \ref{eq:tt_distr}. An optimal algorithm for estimating  $\Delta \Phi$ is reported in  \cite{Bodai2018}. \label{mu098tau}}
\end{figure}


{\color{black}The results given in Eqs. \ref{eq:stationary_distr}-\ref{eq:tt_distr} apply in the case of multiplicative noise laws if one assumes that the noise correlation matrix is well-behaved, according to what discussed in \cite{Lindley2013,Tang2017}.} 
{\color{black}We believe that in our system such a condition applies, essentially because  the factor $1-\alpha(\phi,T_S)$} is {\textit{bounded} in the phase space between 0.4 ($\alpha=\alpha_{max}=0.6$, ice-cover) and 0.8 ($\alpha=\alpha_{min}=0.2$, very warm conditions with  absence of ice cover). In the phase space region near the {\color{black}SB} attractor, we have that $1-\alpha(\phi,T_S)\sim 0.4$ since the temperature $T_S$ is extremely low and the planet is fully glaciated, so that  $\alpha(\phi,T_S)$ is  constant, with $\alpha(\phi,T_S)\sim\alpha_{min}$. 
Near the W attractor, the properties of $\alpha(\phi,T_S)$ are more complex, because only part of the planet is glaciated. {\color{black} If 
$[X]$ the global average of the spatial field $X$, and  $\langle Y\rangle$ is the long term average of $Y$, we have that, typically, $\partial [\langle\alpha\rangle] /\partial  [\langle{T}_S\rangle] <0$, because a decrease in $[\langle T_S\rangle]$ leads to moving the ice line equatorward, thus leading to higher average albedo. 
Then, near the W attractor, noise enhances the instability linked to the {\color{black}$W\rightarrow SB$} transition, and one expects that for finite noise the peak of the  invariant measure is shifted to lower values of $[T_S]$ with respect to the  deterministic attractor.} 


The ratio of the variance of the noise in the {\color{black}W  vs SB attractors can be estimated {\color{black} as} $\approx((1-\alpha_{W})/(1-\alpha_{SB}))^2\approx 3$, where the typical albedo of the W (SB) attractor is $\alpha_{W}\approx0.3$ ($\alpha_{SB}\approx \alpha_{max} =0.6$). T}he two attractors have different microscopic (and macroscopic) temperatures.

We show now our results. We treat two cases inside the region of bistability depicted in Fig. \ref{fig1}, namely $\mu=0.98$ (close to the tipping point $\mu_{W\rightarrow SB}$) and $\mu=1.0$.

In the case of $\mu=0.98$, we consider noise intensities ranging from $\sigma_\tau=0.5\%$ to $\sigma_\tau=1.4\%$, with $\tau=100$ $years$ ($y$). For each value of $\sigma_\tau$,  we initialise 50 orbits in the W basin of attraction and study the statistics of the escape times towards the SB attractor. When the transition takes place, we stop the integration. We observe (not shown) that for each value of $\sigma_\tau$ the escape times are to a good approximation exponentially distributed, see Eq. \ref{eq:tt_distr}. The expectation value of the transition times {\color{black}$\bar{\tau}_\sigma$} is shown in Fig. \ref{mu098tau}. Indeed, {\color{black}$\bar{\tau}_\sigma$} obeys to a  good approximation what shown in Eq. \ref{eq:tt_distr}, so that the difference of the potential {\color{black}$\Phi$} {\color{black}is  half of the} slope of the straight line. For reference, we have that  for $\sigma_{100 y}=0.5\%$ the average escape time is about $5.2\times10^3$ $y$. We can predict that the escape rate increases to about $1.2\times10^7$ $y$ when $\sigma_{100 y}\sim0.3\%$.

We then look at the transition paths. {\color{black}Following \cite{Bodai2014,Lucarini2017}, we choose to consider the reduced phase space spanned by  $[T_S]$ and by $\Delta T_S$, which is the difference between the spatial averages of $T_S$ in the latitudinal belts $[0,30 ^\circ N]$ and $[30 ^\circ N,90 ^\circ N]$, respectively.} This reduced phase space provides a minimal yet physically informative viewpoint on the problem. Figure \ref{mu098} depicts ($\sigma_\tau=1.0\%$), the {\color{black}transient two-dimensional probability distribution function (pdf) $\tilde\rho$ constructed using the above-described 50 simulations, where  the statistics is collected only until the {\color{black}$W\rightarrow SB$} transition is realised. \color{black}Note that $\tilde\rho$ {\color{black} is \textit{not}} the invariant density of the system.} The transitions {\color{black} typically take} place along a very narrow band linking the W attractor  and the M state 
\footnote{The W attractor and the M state are not dots, as they are chaotic (see Fig. \ref{fig1}), but they have small variability in the projected space $([T_S],\Delta T_S)$.}. 
We  construct an estimate of the instanton associated to the {\color{black}$W\rightarrow SB$} transition by conditionally averaging the orbits according to the value of $[T_S]$. To a good approximation, the instanton connects the W attractor to the {\color{black}M} state, and follows a path of decreasing probability.  {\color{black}We remark that we} do not find evidence of different paths for  escape vs relaxation trajectories, which, instead, is a signature of non-equilibrium \cite{ZinnJustin1996}. This can be explained by considering {\color{black} that, as discussed in \cite{Bodai2014}}, the ocean model evolve{\color{black}s} approximately in an energy landscape.

\begin{figure}
\includegraphics[trim=1cm 0.5cm 1cm 1.9cm, clip=true, width=0.47\textwidth]{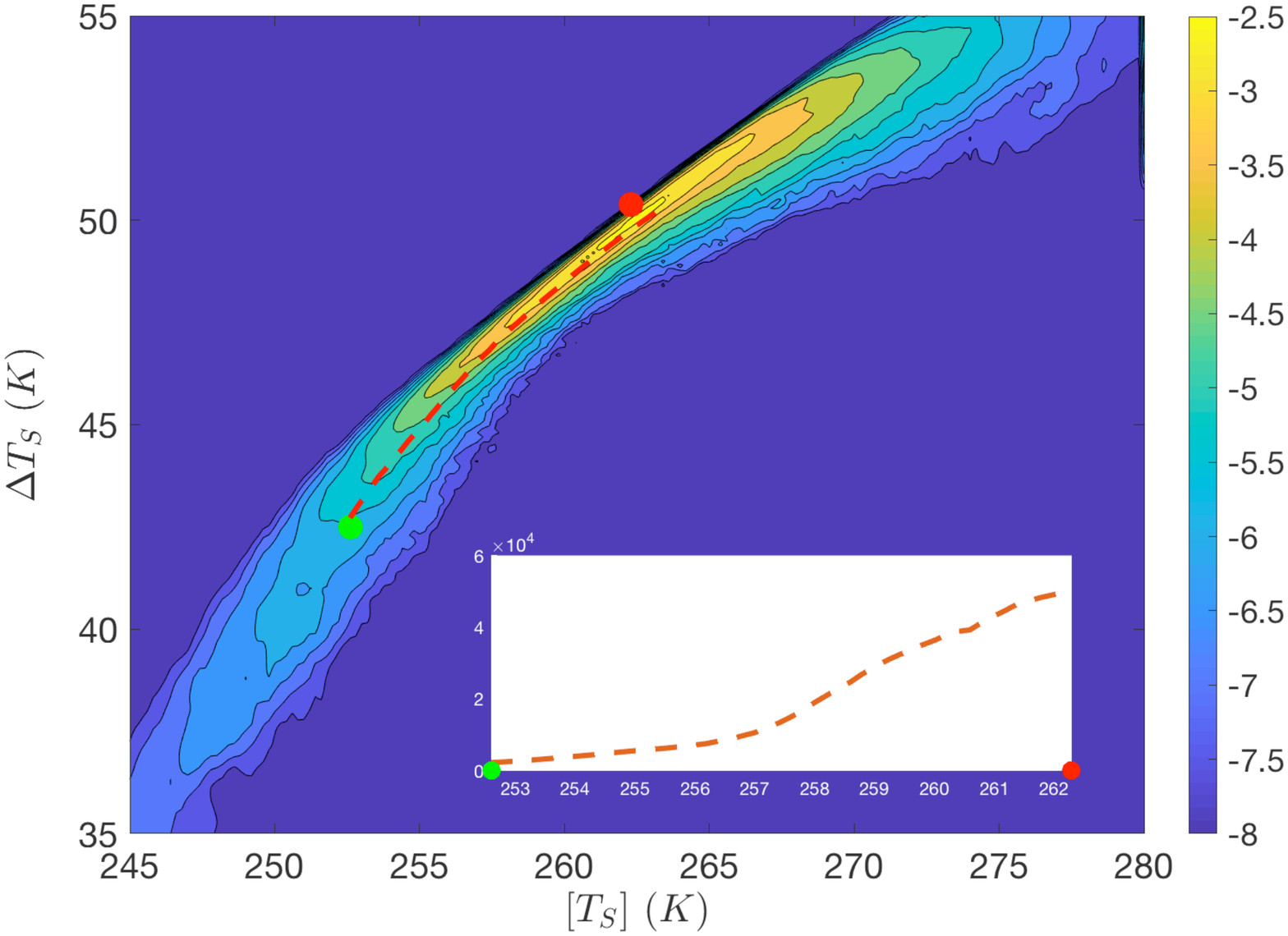}%
\caption{Main graph: Logarithm of $\tilde{\rho}$ projected onto $(T_S,[\Delta T_S])$ for $\mu=0.98$; W attractor {\color{black}(red dot); {\color{black}M} state (green dot)}. We  have used $\sigma_{100y}=1\%$. The {\color{black}$W\rightarrow SB$} instanton (red dashed line) is indicated. {\color{black}Bottom} right inset: probability along the instanton. \label{mu098}}
\end{figure}


{\color{black}When $\mu=0.98$, the {\color{black}$SB\rightarrow W$} transitions are rather rare unless one considers relatively large values of $\sigma$. This is due to the much lower value of $\Phi$ at the SB attractor than at the W attractor (see Eq. \ref{eq:tt_distr}), so that (see Eq. \ref{eq:stationary_distr}) the fraction of the invariant measure supported near the W attractor is extremely small. 
We focus next on the case of $\mu=1.0$, where 
the population is split more evenly between the W and SB attractors.}  
We are {\color{black} then able to construct for each value of $\sigma_\tau$ the invariant measure using a single orbit of the system, provided we can observe a sufficient number of transitions.  We   use $\sigma_{100y}=1.5\%$. Our results are shown in Fig. \ref{lownoise} for a trajectory lasting $\approx6.0\times10^4$ $y$ and characterised by 92 {\color{black}$SB\rightarrow W$} and {\color{black}$W\rightarrow SB$} transitions, whose average rates are consistent with an occupation of about $35\%$ for the $W$ basin of attraction, and of about $65\%$ for the {\color{black}$SB$} basin of attraction. The projection  of the invariant measure on the $([T_S],\Delta T_S)$ plane shows that the peaks of the pdfs are very close to the $W$ and $SB$ attractors {\color{black}(note the predicted slight shift for the $W$ case, visible because the noise is stronger than in Fig. \ref{mu098})}, and that the agreement further improves when considering the two marginal {\color{black}pdfs} (top left and bottom right insets). We can construct both the {\color{black}$W\rightarrow SB$} and the {\color{black}$SB\rightarrow W$} instantons, whose starting and final points agree remarkably well with the attractors and {\color{black}the M} state. We discover that the instantons  follow a path of monotonic descent{\color{black}, following} closely the crests of the pdf), with the minimum at the M state}. 
We also run a  simulation lasting $\approx2.7\times10^4$ $y$ using $\sigma_{100y}=1.8\%$. We obtain 73 {\color{black}$SB\rightarrow W$} and {\color{black}$W\rightarrow SB$} transitions. Comparing the statistics of this run with what is shown in Fig. \ref{lownoise}, we find good agreement with the predictions of  Eqs. \ref{eq:stationary_distr}-\ref{eq:tt_distr} (not shown).

\begin{figure}
\includegraphics[trim=2cm .5cm 1cm 1.9cm, clip=true, width=0.47\textwidth]{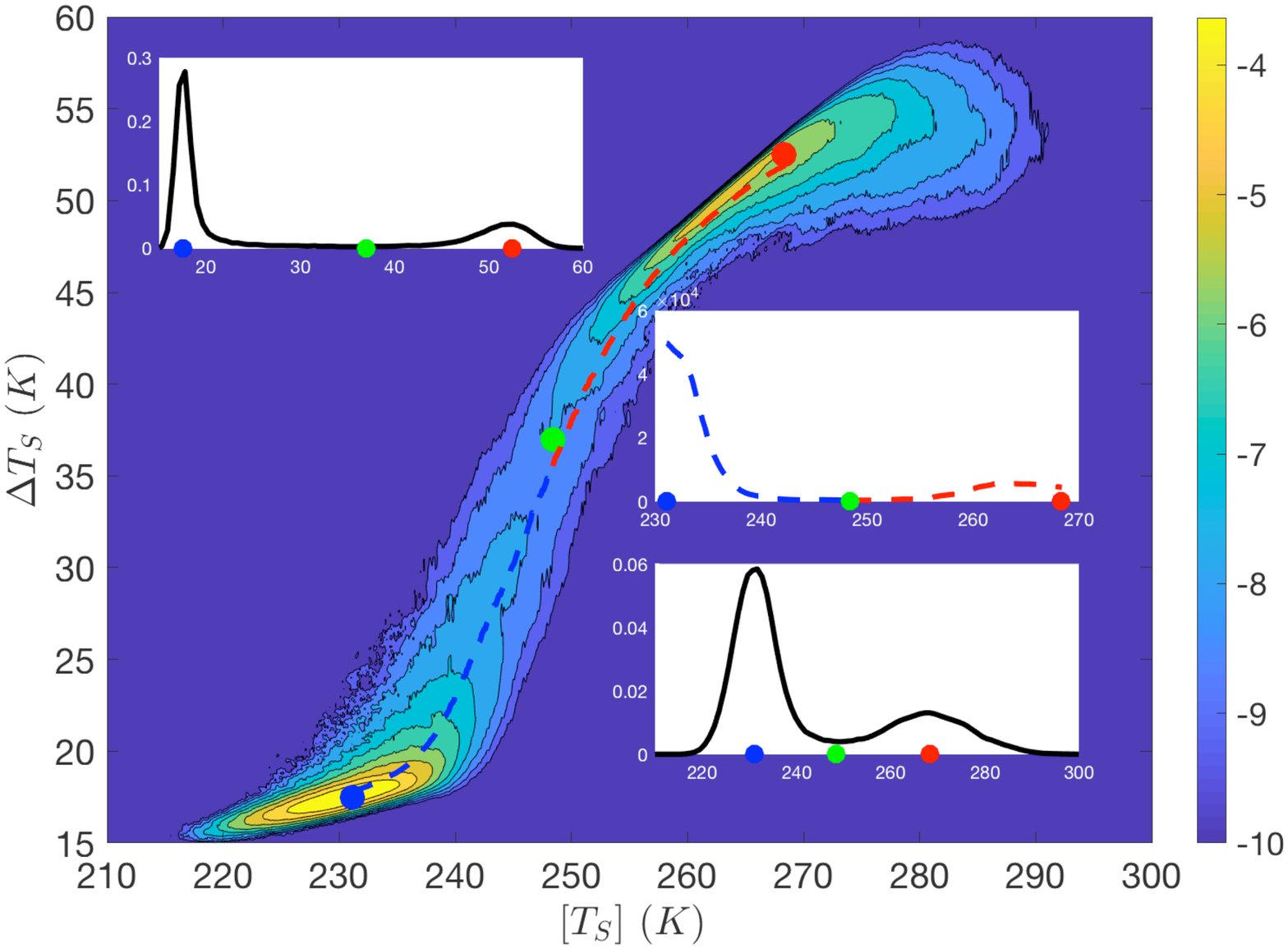}
\caption{Main graph: $\rho$  in the projected phase space $(T_S,[\Delta T_S])$. {\color{black}W attractor (red dot), SB attractor (blue dot), M state (green dot) }for $\mu=1$. Red (blue) dashed line: {\color{black}$W\rightarrow SB$} ({\color{black}$SB\rightarrow W$}) instanton. We  have used $\sigma_{100y}=1.5\%$. Top left inset: marginal pdf with respect to $\Delta T_S$. Bottom right inset: marginal pdf with respect to $[T_S]$.  Center {\color{black}right} inset: probability along the two instantons. \label{lownoise}}
\end{figure}

Concluding, in this {\color{black}letter} we have studied the problem of noise-induced transitions between the W and the SB state of the climate system using a intermediate complexity stochastic model. The deterministic version of this model had been  used to construct the M states for a vast range of values of the  solar irradiance \cite{Lucarini2017}. {\color{black} Including stochastic perturbations allows for exploring the phase space of the system. We have chosen to study the impact of fluctuations in the  solar irradiance, which entails adding a multiplicative noise. In particular, the SB climate reflects more radiation because large ice-covered surfaces lead to higher albedo, so that the effective noise will be weaker than the one acting near the W attractor.  We have explained why a large deviation theory-based mathematical framework is able to explain very satisfactorily our results}. We show that one can construct the {\color{black}pseudo-}potential defining the escape rate from the basins of attraction and the natural measure of the stochastically perturbed system.  {\color{black} In future investigations we plan to extend the analysis performed here to values of $\mu$ covering the whole range of bistability, in order to understanding  how the pseudo-potential depends on $\mu$.} Using a low-dimensional projection of the invariant measure, we  show that  the instantons connect the attractors with the M states, following a path of steepest descent. This gives a key connection between the stochastic and the deterministic points of view on study of the multistability of the climate. 

A further link between the stochastically perturbed and deterministic system can be described as follows. If we consider values of $\mu$ just below the critical one {\color{black}$\mu_{W\rightarrow SB}\sim 0.965$} defining the {\color{black}$W\rightarrow SB$} tipping point, one can find long-lived transient chaotic trajectories. {\color{black}It is worth investigating whether such transients are associated to a invariant saddle emerging as a result of} the boundary crisis. 
We have observed that the way such long-lived trajectories collapse to the SB state is very similar to the way, when $\mu=0.98$, stochastically perturbed orbits initialised in the W state basin of attraction  perform the transition. {\color{black}In future studies we will explore such similarity 
looking into the possible presence of a \textit{ghost state} \cite{Medeiros2017}}.

{\color{black} The knowledge of M states is key to understanding tipping points. When an attractor and a M states are nearby, the system's response to perturbations diverges, as the correlations decay very slowly \cite{Tantet2018}. These findings extend and generalise the classical point of view given in \cite{Lenton2008}. What is proposed here, combined with the framework given in \cite{Ashwin2012}, could be key for understanding and predicting the criticalities in the trajectories of the Earth system \cite{Steffen2018}, including those leading to very hot, non-habitable conditions \cite{Gomez2018}. A tipping element we will investigate along the lines of the present study is the Atlantic meridional overturning circulation \cite{Lenton2008}}.

\section*{Acknowledgments}
{\color{black}The authors wish to thank G. Drotos and J. Wouters for useful exchanges, }and  acknowledge the support received by the EU Horizon2020 projects Blue-Action (grant No. 727852) and CRESCENDO (grant No. 641816). VL acknowledges the support of the DFG SFB/Transregio project TRR181. 

\begin{thebibliography}{41}%
\makeatletter
\providecommand \@ifxundefined [1]{%
 \@ifx{#1\undefined}
}%
\providecommand \@ifnum [1]{%
 \ifnum #1\expandafter \@firstoftwo
 \else \expandafter \@secondoftwo
 \fi
}%
\providecommand \@ifx [1]{%
 \ifx #1\expandafter \@firstoftwo
 \else \expandafter \@secondoftwo
 \fi
}%
\providecommand \natexlab [1]{#1}%
\providecommand \enquote  [1]{``#1''}%
\providecommand \bibnamefont  [1]{#1}%
\providecommand \bibfnamefont [1]{#1}%
\providecommand \citenamefont [1]{#1}%
\providecommand \href@noop [0]{\@secondoftwo}%
\providecommand \href [0]{\begingroup \@sanitize@url \@href}%
\providecommand \@href[1]{\@@startlink{#1}\@@href}%
\providecommand \@@href[1]{\endgroup#1\@@endlink}%
\providecommand \@sanitize@url [0]{\catcode `\\12\catcode `\$12\catcode
  `\&12\catcode `\#12\catcode `\^12\catcode `\_12\catcode `\%12\relax}%
\providecommand \@@startlink[1]{}%
\providecommand \@@endlink[0]{}%
\providecommand \url  [0]{\begingroup\@sanitize@url \@url }%
\providecommand \@url [1]{\endgroup\@href {#1}{\urlprefix }}%
\providecommand \urlprefix  [0]{URL }%
\providecommand \Eprint [0]{\href }%
\providecommand \doibase [0]{http://dx.doi.org/}%
\providecommand \selectlanguage [0]{\@gobble}%
\providecommand \bibinfo  [0]{\@secondoftwo}%
\providecommand \bibfield  [0]{\@secondoftwo}%
\providecommand \translation [1]{[#1]}%
\providecommand \BibitemOpen [0]{}%
\providecommand \bibitemStop [0]{}%
\providecommand \bibitemNoStop [0]{.\EOS\space}%
\providecommand \EOS [0]{\spacefactor3000\relax}%
\providecommand \BibitemShut  [1]{\csname bibitem#1\endcsname}%
\let\auto@bib@innerbib\@empty
\bibitem [{\citenamefont {Hoffman}\ and\ \citenamefont
  {Schrag}(2002)}]{Hoffman2002}%
  \BibitemOpen
  \bibfield  {author} {\bibinfo {author} {\bibfnamefont {P.~F.}\ \bibnamefont
  {Hoffman}}\ and\ \bibinfo {author} {\bibfnamefont {D.~P.}\ \bibnamefont
  {Schrag}},\ }\href {\doibase 10.1046/j.1365-3121.2002.00408.x} {\bibfield
  {journal} {\bibinfo  {journal} {Terra Nova}\ }\textbf {\bibinfo {volume}
  {14}},\ \bibinfo {pages} {129} (\bibinfo {year} {2002})}\BibitemShut
  {NoStop}%
\bibitem [{\citenamefont {Pierrehumbert}\ \emph {et~al.}(2011)\citenamefont
  {Pierrehumbert}, \citenamefont {Abbot}, \citenamefont {Voigt},\ and\
  \citenamefont {Koll}}]{Pierrehumbert2011}%
  \BibitemOpen
  \bibfield  {author} {\bibinfo {author} {\bibfnamefont {R.~T.}\ \bibnamefont
  {Pierrehumbert}}, \bibinfo {author} {\bibfnamefont {D.}~\bibnamefont
  {Abbot}}, \bibinfo {author} {\bibfnamefont {A.}~\bibnamefont {Voigt}}, \ and\
  \bibinfo {author} {\bibfnamefont {D.}~\bibnamefont {Koll}},\ }\href@noop {}
  {\bibfield  {journal} {\bibinfo  {journal} {Ann. Rev, Earth Plan. Sci.}\
  }\textbf {\bibinfo {volume} {39}},\ \bibinfo {pages} {417} (\bibinfo {year}
  {2011})}\BibitemShut {NoStop}%
\bibitem [{\citenamefont {Donnadieu}\ \emph {et~al.}(2014)\citenamefont
  {Donnadieu}, \citenamefont {Goddéris},\ and\ \citenamefont
  {Hir}}]{donnadieu2014}%
  \BibitemOpen
  \bibfield  {author} {\bibinfo {author} {\bibfnamefont {Y.}~\bibnamefont
  {Donnadieu}}, \bibinfo {author} {\bibfnamefont {Y.}~\bibnamefont
  {Goddéris}}, \ and\ \bibinfo {author} {\bibfnamefont {G.~L.}\ \bibnamefont
  {Hir}},\ }in\ \href {\doibase
  https://doi.org/10.1016/B978-0-08-095975-7.01309-7} {\emph {\bibinfo
  {booktitle} {Treatise on Geochemistry}}},\ \bibinfo {editor} {edited by\
  \bibinfo {editor} {\bibfnamefont {H.~D.}\ \bibnamefont {Holland}}\ and\
  \bibinfo {editor} {\bibfnamefont {K.~K.}\ \bibnamefont {Turekian}}}\
  (\bibinfo  {publisher} {Elsevier},\ \bibinfo {address} {Oxford},\ \bibinfo
  {year} {2014})\ pp.\ \bibinfo {pages} {217 -- 229}\BibitemShut {NoStop}%
\bibitem [{\citenamefont {Budyko}(1969)}]{Budyko1969}%
  \BibitemOpen
  \bibfield  {author} {\bibinfo {author} {\bibfnamefont {M.}~\bibnamefont
  {Budyko}},\ }\href@noop {} {\bibfield  {journal} {\bibinfo  {journal}
  {Tellus}\ }\textbf {\bibinfo {volume} {21}},\ \bibinfo {pages} {611}
  (\bibinfo {year} {1969})}\BibitemShut {NoStop}%
\bibitem [{\citenamefont {Sellers}(1969)}]{Sellers1969}%
  \BibitemOpen
  \bibfield  {author} {\bibinfo {author} {\bibfnamefont {W.}~\bibnamefont
  {Sellers}},\ }\href@noop {} {\bibfield  {journal} {\bibinfo  {journal} {J.
  Appl. Meteorol.}\ }\textbf {\bibinfo {volume} {8}},\ \bibinfo {pages} {392}
  (\bibinfo {year} {1969})}\BibitemShut {NoStop}%
\bibitem [{\citenamefont {Ghil}(1976)}]{Ghil1976}%
  \BibitemOpen
  \bibfield  {author} {\bibinfo {author} {\bibfnamefont {M.}~\bibnamefont
  {Ghil}},\ }\href@noop {} {\bibfield  {journal} {\bibinfo  {journal} {J.
  Atmos. Sci.}\ }\textbf {\bibinfo {volume} {33}},\ \bibinfo {pages} {3}
  (\bibinfo {year} {1976})}\BibitemShut {NoStop}%
\bibitem [{\citenamefont {Lucarini}\ \emph {et~al.}(2013)\citenamefont
  {Lucarini}, \citenamefont {Pascale}, \citenamefont {Boschi}, \citenamefont
  {Kirk},\ and\ \citenamefont {Iro}}]{LucAstr2013}%
  \BibitemOpen
  \bibfield  {author} {\bibinfo {author} {\bibfnamefont {V.}~\bibnamefont
  {Lucarini}}, \bibinfo {author} {\bibfnamefont {S.}~\bibnamefont {Pascale}},
  \bibinfo {author} {\bibfnamefont {R.}~\bibnamefont {Boschi}}, \bibinfo
  {author} {\bibfnamefont {E.}~\bibnamefont {Kirk}}, \ and\ \bibinfo {author}
  {\bibfnamefont {N.}~\bibnamefont {Iro}},\ }\href@noop {} {\bibfield
  {journal} {\bibinfo  {journal} {Astr. Nach.}\ }\textbf {\bibinfo {volume}
  {334}},\ \bibinfo {pages} {576} (\bibinfo {year} {2013})}\BibitemShut
  {NoStop}%
\bibitem [{\citenamefont {Boschi}\ \emph {et~al.}(2013)\citenamefont {Boschi},
  \citenamefont {Lucarini},\ and\ \citenamefont {Pascale}}]{Boschi2013}%
  \BibitemOpen
  \bibfield  {author} {\bibinfo {author} {\bibfnamefont {R.}~\bibnamefont
  {Boschi}}, \bibinfo {author} {\bibfnamefont {V.}~\bibnamefont {Lucarini}}, \
  and\ \bibinfo {author} {\bibfnamefont {S.}~\bibnamefont {Pascale}},\ }\href
  {\doibase http://dx.doi.org/10.1016/j.icarus.2013.03.017} {\bibfield
  {journal} {\bibinfo  {journal} {Icarus}\ }\textbf {\bibinfo {volume} {227}},\
  \bibinfo {pages} {1724} (\bibinfo {year} {2013})}\BibitemShut {NoStop}%
\bibitem [{\citenamefont {Lucarini}\ and\ \citenamefont
  {B\'odai}(2017)}]{Lucarini2017}%
  \BibitemOpen
  \bibfield  {author} {\bibinfo {author} {\bibfnamefont {V.}~\bibnamefont
  {Lucarini}}\ and\ \bibinfo {author} {\bibfnamefont {T.}~\bibnamefont
  {B\'odai}},\ }\href {http://stacks.iop.org/0951-7715/30/i=7/a=R32} {\bibfield
   {journal} {\bibinfo  {journal} {Nonlinearity}\ }\textbf {\bibinfo {volume}
  {30}},\ \bibinfo {pages} {R32} (\bibinfo {year} {2017})}\BibitemShut
  {NoStop}%
\bibitem [{\citenamefont {Feudel}\ \emph {et~al.}(2018)\citenamefont {Feudel},
  \citenamefont {Pisarchik},\ and\ \citenamefont {Showalter}}]{Feudel2018}%
  \BibitemOpen
  \bibfield  {author} {\bibinfo {author} {\bibfnamefont {U.}~\bibnamefont
  {Feudel}}, \bibinfo {author} {\bibfnamefont {A.~N.}\ \bibnamefont
  {Pisarchik}}, \ and\ \bibinfo {author} {\bibfnamefont {K.}~\bibnamefont
  {Showalter}},\ }\href {\doibase 10.1063/1.5027718} {\bibfield  {journal}
  {\bibinfo  {journal} {Chaos: An Interdisciplinary Journal of Nonlinear
  Science}\ }\textbf {\bibinfo {volume} {28}},\ \bibinfo {pages} {033501}
  (\bibinfo {year} {2018})}\BibitemShut {NoStop}%
\bibitem [{\citenamefont {Lenton}\ and\ \citenamefont
  {et~al.}(2008)}]{Lenton2008}%
  \BibitemOpen
  \bibfield  {author} {\bibinfo {author} {\bibfnamefont {T.}~\bibnamefont
  {Lenton}}\ and\ \bibinfo {author} {\bibnamefont {et~al.}},\ }\href {\doibase
  10.1073/pnas.0705414105} {\bibfield  {journal} {\bibinfo  {journal}
  {Proceedings of the National Academy of Sciences}\ }\textbf {\bibinfo
  {volume} {105}},\ \bibinfo {pages} {1786} (\bibinfo {year}
  {2008})}\BibitemShut {NoStop}%
\bibitem [{\citenamefont {Grebogi}\ \emph {et~al.}(1983)\citenamefont
  {Grebogi}, \citenamefont {Ott},\ and\ \citenamefont {Yorke}}]{Grebogi1983}%
  \BibitemOpen
  \bibfield  {author} {\bibinfo {author} {\bibfnamefont {C.}~\bibnamefont
  {Grebogi}}, \bibinfo {author} {\bibfnamefont {E.}~\bibnamefont {Ott}}, \ and\
  \bibinfo {author} {\bibfnamefont {J.~A.}\ \bibnamefont {Yorke}},\ }\href
  {\doibase 10.1103/PhysRevLett.50.935} {\bibfield  {journal} {\bibinfo
  {journal} {Phys. Rev. Lett.}\ }\textbf {\bibinfo {volume} {50}},\ \bibinfo
  {pages} {935} (\bibinfo {year} {1983})}\BibitemShut {NoStop}%
\bibitem [{\citenamefont {Robert}\ \emph {et~al.}(2000)\citenamefont {Robert},
  \citenamefont {Alligood}, \citenamefont {Ott},\ and\ \citenamefont
  {Yorke}}]{Robert2000}%
  \BibitemOpen
  \bibfield  {author} {\bibinfo {author} {\bibfnamefont {C.}~\bibnamefont
  {Robert}}, \bibinfo {author} {\bibfnamefont {K.~T.}\ \bibnamefont
  {Alligood}}, \bibinfo {author} {\bibfnamefont {E.}~\bibnamefont {Ott}}, \
  and\ \bibinfo {author} {\bibfnamefont {J.~A.}\ \bibnamefont {Yorke}},\ }\href
  {\doibase https://doi.org/10.1016/S0167-2789(00)00074-9} {\bibfield
  {journal} {\bibinfo  {journal} {Physica D: Nonlinear Phenomena}\ }\textbf
  {\bibinfo {volume} {144}},\ \bibinfo {pages} {44 } (\bibinfo {year}
  {2000})}\BibitemShut {NoStop}%
\bibitem [{\citenamefont {Ott}(2002)}]{Ott2002}%
  \BibitemOpen
  \bibfield  {author} {\bibinfo {author} {\bibfnamefont {E.}~\bibnamefont
  {Ott}},\ }\href {https://books.google.co.uk/books?id=nOLx--zzHSgC} {\emph
  {\bibinfo {title} {Chaos in Dynamical Systems}}}\ (\bibinfo  {publisher}
  {Cambridge University Press},\ \bibinfo {year} {2002})\BibitemShut {NoStop}%
\bibitem [{\citenamefont {Vollmer}\ \emph {et~al.}(2009)\citenamefont
  {Vollmer}, \citenamefont {Schneider},\ and\ \citenamefont
  {Eckhardt}}]{Vollmer2009}%
  \BibitemOpen
  \bibfield  {author} {\bibinfo {author} {\bibfnamefont {J.}~\bibnamefont
  {Vollmer}}, \bibinfo {author} {\bibfnamefont {T.~M.}\ \bibnamefont
  {Schneider}}, \ and\ \bibinfo {author} {\bibfnamefont {B.}~\bibnamefont
  {Eckhardt}},\ }\href {http://stacks.iop.org/1367-2630/11/i=1/a=013040}
  {\bibfield  {journal} {\bibinfo  {journal} {New Journal of Physics}\ }\textbf
  {\bibinfo {volume} {11}},\ \bibinfo {pages} {013040} (\bibinfo {year}
  {2009})}\BibitemShut {NoStop}%
\bibitem [{\citenamefont {B{\'o}dai}\ \emph {et~al.}(2014)\citenamefont
  {B{\'o}dai}, \citenamefont {Lucarini}, \citenamefont {Lunkeit},\ and\
  \citenamefont {Boschi}}]{Bodai2014}%
  \BibitemOpen
  \bibfield  {author} {\bibinfo {author} {\bibfnamefont {T.}~\bibnamefont
  {B{\'o}dai}}, \bibinfo {author} {\bibfnamefont {V.}~\bibnamefont {Lucarini}},
  \bibinfo {author} {\bibfnamefont {F.}~\bibnamefont {Lunkeit}}, \ and\
  \bibinfo {author} {\bibfnamefont {R.}~\bibnamefont {Boschi}},\ }\href
  {\doibase 10.1007/s00382-014-2206-5} {\bibfield  {journal} {\bibinfo
  {journal} {Clim. Dyn.}\ }\textbf {\bibinfo {volume} {44}},\ \bibinfo {pages}
  {3361} (\bibinfo {year} {2014})}\BibitemShut {NoStop}%
\bibitem [{\citenamefont {Skufca}\ \emph {et~al.}(2006)\citenamefont {Skufca},
  \citenamefont {Yorke},\ and\ \citenamefont {Eckhardt}}]{Skufca2006}%
  \BibitemOpen
  \bibfield  {author} {\bibinfo {author} {\bibfnamefont {J.~D.}\ \bibnamefont
  {Skufca}}, \bibinfo {author} {\bibfnamefont {J.~A.}\ \bibnamefont {Yorke}}, \
  and\ \bibinfo {author} {\bibfnamefont {B.}~\bibnamefont {Eckhardt}},\ }\href
  {\doibase 10.1103/PhysRevLett.96.174101} {\bibfield  {journal} {\bibinfo
  {journal} {Phys. Rev. Lett.}\ }\textbf {\bibinfo {volume} {96}},\ \bibinfo
  {pages} {174101} (\bibinfo {year} {2006})}\BibitemShut {NoStop}%
\bibitem [{\citenamefont {Schneider}\ \emph {et~al.}(2007)\citenamefont
  {Schneider}, \citenamefont {Eckhardt},\ and\ \citenamefont
  {Yorke}}]{Schneider2007}%
  \BibitemOpen
  \bibfield  {author} {\bibinfo {author} {\bibfnamefont {T.~M.}\ \bibnamefont
  {Schneider}}, \bibinfo {author} {\bibfnamefont {B.}~\bibnamefont {Eckhardt}},
  \ and\ \bibinfo {author} {\bibfnamefont {J.~A.}\ \bibnamefont {Yorke}},\
  }\href {\doibase 10.1103/PhysRevLett.99.034502} {\bibfield  {journal}
  {\bibinfo  {journal} {Phys. Rev. Lett.}\ }\textbf {\bibinfo {volume} {99}},\
  \bibinfo {pages} {034502} (\bibinfo {year} {2007})}\BibitemShut {NoStop}%
\bibitem [{\citenamefont {Tantet}\ \emph {et~al.}(2018)\citenamefont {Tantet},
  \citenamefont {Lucarini}, \citenamefont {Lunkeit},\ and\ \citenamefont
  {Dijkstra}}]{Tantet2018}%
  \BibitemOpen
  \bibfield  {author} {\bibinfo {author} {\bibfnamefont {A.}~\bibnamefont
  {Tantet}}, \bibinfo {author} {\bibfnamefont {V.}~\bibnamefont {Lucarini}},
  \bibinfo {author} {\bibfnamefont {F.}~\bibnamefont {Lunkeit}}, \ and\
  \bibinfo {author} {\bibfnamefont {H.~A.}\ \bibnamefont {Dijkstra}},\ }\href
  {http://stacks.iop.org/0951-7715/31/i=5/a=2221} {\bibfield  {journal}
  {\bibinfo  {journal} {Nonlinearity}\ }\textbf {\bibinfo {volume} {31}},\
  \bibinfo {pages} {2221} (\bibinfo {year} {2018})}\BibitemShut {NoStop}%
\bibitem [{\citenamefont {Lai}\ and\ \citenamefont {T\'el}(2011)}]{LT:2011}%
  \BibitemOpen
  \bibfield  {author} {\bibinfo {author} {\bibfnamefont {Y.-C.}\ \bibnamefont
  {Lai}}\ and\ \bibinfo {author} {\bibfnamefont {T.}~\bibnamefont {T\'el}},\
  }\href@noop {} {\emph {\bibinfo {title} {Transient Chaos}}}\ (\bibinfo
  {publisher} {Springer},\ \bibinfo {address} {New York},\ \bibinfo {year}
  {2011})\BibitemShut {NoStop}%
\bibitem [{\citenamefont {Lorenz}(1975)}]{Lorenz1975}%
  \BibitemOpen
  \bibfield  {author} {\bibinfo {author} {\bibfnamefont {E.~N.}\ \bibnamefont
  {Lorenz}},\ }in\ \href@noop {} {\emph {\bibinfo {booktitle} {GARP Publication
  Series}}}\ (\bibinfo  {publisher} {WMO},\ \bibinfo {year} {1975})\ pp.\
  \bibinfo {pages} {132--136}\BibitemShut {NoStop}%
\bibitem [{\citenamefont {Frisius}\ \emph {et~al.}(1998)\citenamefont
  {Frisius}, \citenamefont {Lunkeit}, \citenamefont {Fraedrich},\ and\
  \citenamefont {James}}]{puma}%
  \BibitemOpen
  \bibfield  {author} {\bibinfo {author} {\bibfnamefont {T.}~\bibnamefont
  {Frisius}}, \bibinfo {author} {\bibfnamefont {F.}~\bibnamefont {Lunkeit}},
  \bibinfo {author} {\bibfnamefont {K.}~\bibnamefont {Fraedrich}}, \ and\
  \bibinfo {author} {\bibfnamefont {I.~N.}\ \bibnamefont {James}},\ }\href
  {\doibase 10.1002/qj.49712454802} {\bibfield  {journal} {\bibinfo  {journal}
  {Quarterly Journal of the Royal Meteorological Society}\ }\textbf {\bibinfo
  {volume} {124}},\ \bibinfo {pages} {1019} (\bibinfo {year}
  {1998})}\BibitemShut {NoStop}%
\bibitem [{\citenamefont {Schneider}\ and\ \citenamefont
  {Gal-Chen}(1973)}]{Schneider1973}%
  \BibitemOpen
  \bibfield  {author} {\bibinfo {author} {\bibfnamefont {S.~H.}\ \bibnamefont
  {Schneider}}\ and\ \bibinfo {author} {\bibfnamefont {T.}~\bibnamefont
  {Gal-Chen}},\ }\href {\doibase 10.1029/JC078i027p06182} {\bibfield  {journal}
  {\bibinfo  {journal} {Journal of Geophysical Research}\ }\textbf {\bibinfo
  {volume} {78}},\ \bibinfo {pages} {6182} (\bibinfo {year}
  {1973})}\BibitemShut {NoStop}%
\bibitem [{\citenamefont {Dwyer}\ and\ \citenamefont
  {Petersen}(1973)}]{Dwyer1973}%
  \BibitemOpen
  \bibfield  {author} {\bibinfo {author} {\bibfnamefont {H.~A.}\ \bibnamefont
  {Dwyer}}\ and\ \bibinfo {author} {\bibfnamefont {T.}~\bibnamefont
  {Petersen}},\ }\href {\doibase
  10.1175/1520-0450(1973)012<0036:TDGEM>2.0.CO;2} {\bibfield  {journal}
  {\bibinfo  {journal} {J. Appl. Meteo.}\ }\textbf {\bibinfo {volume} {12}},\
  \bibinfo {pages} {36} (\bibinfo {year} {1973})}\BibitemShut {NoStop}%
\bibitem [{\citenamefont {Saltzman}(2001)}]{saltzman_dynamical}%
  \BibitemOpen
  \bibfield  {author} {\bibinfo {author} {\bibfnamefont {B.}~\bibnamefont
  {Saltzman}},\ }\href@noop {} {\emph {\bibinfo {title} {Dynamical
  Paleoclimatology}}}\ (\bibinfo  {publisher} {Academic Press New York},\
  \bibinfo {year} {2001})\BibitemShut {NoStop}%
\bibitem [{\citenamefont {Hanggi}(1986)}]{Hanggi1986}%
  \BibitemOpen
  \bibfield  {author} {\bibinfo {author} {\bibfnamefont {P.}~\bibnamefont
  {Hanggi}},\ }\href {\doibase 10.1007/BF01010843} {\bibfield  {journal}
  {\bibinfo  {journal} {Journal of Statistical Physics}\ }\textbf {\bibinfo
  {volume} {42}},\ \bibinfo {pages} {105} (\bibinfo {year} {1986})}\BibitemShut
  {NoStop}%
\bibitem [{\citenamefont {Kautz}(1987)}]{Kautz1987}%
  \BibitemOpen
  \bibfield  {author} {\bibinfo {author} {\bibfnamefont {R.}~\bibnamefont
  {Kautz}},\ }\href {\doibase https://doi.org/10.1016/0375-9601(87)90151-4}
  {\bibfield  {journal} {\bibinfo  {journal} {Physics Letters A}\ }\textbf
  {\bibinfo {volume} {125}},\ \bibinfo {pages} {315 } (\bibinfo {year}
  {1987})}\BibitemShut {NoStop}%
\bibitem [{\citenamefont {Grassberger}(1989)}]{Grassberger1989}%
  \BibitemOpen
  \bibfield  {author} {\bibinfo {author} {\bibfnamefont {P.}~\bibnamefont
  {Grassberger}},\ }\href {http://stacks.iop.org/0305-4470/22/i=16/a=018}
  {\bibfield  {journal} {\bibinfo  {journal} {J. Phys. A}\ }\textbf {\bibinfo
  {volume} {22}},\ \bibinfo {pages} {3283} (\bibinfo {year}
  {1989})}\BibitemShut {NoStop}%
\bibitem [{\citenamefont {Freidlin}\ and\ \citenamefont
  {Wentzell}(1984)}]{Freidlin1984}%
  \BibitemOpen
  \bibfield  {author} {\bibinfo {author} {\bibfnamefont {M.~I.}\ \bibnamefont
  {Freidlin}}\ and\ \bibinfo {author} {\bibfnamefont {A.}~\bibnamefont
  {Wentzell}},\ }\href@noop {} {\emph {\bibinfo {title} {Random Perturbations
  of Dynamical Systems}}}\ (\bibinfo  {publisher} {Springer},\ \bibinfo
  {address} {New York},\ \bibinfo {year} {1984})\BibitemShut {NoStop}%
\bibitem [{\citenamefont {Graham}\ \emph {et~al.}(1991)\citenamefont {Graham},
  \citenamefont {Hamm},\ and\ \citenamefont {T\'el}}]{Graham1991}%
  \BibitemOpen
  \bibfield  {author} {\bibinfo {author} {\bibfnamefont {R.}~\bibnamefont
  {Graham}}, \bibinfo {author} {\bibfnamefont {A.}~\bibnamefont {Hamm}}, \ and\
  \bibinfo {author} {\bibfnamefont {T.}~\bibnamefont {T\'el}},\ }\href
  {\doibase 10.1103/PhysRevLett.66.3089} {\bibfield  {journal} {\bibinfo
  {journal} {Phys. Rev. Lett.}\ }\textbf {\bibinfo {volume} {66}},\ \bibinfo
  {pages} {3089} (\bibinfo {year} {1991})}\BibitemShut {NoStop}%
\bibitem [{\citenamefont {Hamm}\ \emph {et~al.}(1994)\citenamefont {Hamm},
  \citenamefont {T\'el},\ and\ \citenamefont {Graham}}]{Hamm1994}%
  \BibitemOpen
  \bibfield  {author} {\bibinfo {author} {\bibfnamefont {A.}~\bibnamefont
  {Hamm}}, \bibinfo {author} {\bibfnamefont {T.}~\bibnamefont {T\'el}}, \ and\
  \bibinfo {author} {\bibfnamefont {R.}~\bibnamefont {Graham}},\ }\href
  {\doibase https://doi.org/10.1016/0375-9601(94)90621-1} {\bibfield  {journal}
  {\bibinfo  {journal} {Phys. Lett. A}\ }\textbf {\bibinfo {volume} {185}},\
  \bibinfo {pages} {313 } (\bibinfo {year} {1994})}\BibitemShut {NoStop}%
\bibitem [{\citenamefont {Kraut}\ and\ \citenamefont
  {Feudel}(2002)}]{Kraut2002}%
  \BibitemOpen
  \bibfield  {author} {\bibinfo {author} {\bibfnamefont {S.}~\bibnamefont
  {Kraut}}\ and\ \bibinfo {author} {\bibfnamefont {U.}~\bibnamefont {Feudel}},\
  }\href {\doibase 10.1103/PhysRevE.66.015207} {\bibfield  {journal} {\bibinfo
  {journal} {Phys. Rev. E}\ }\textbf {\bibinfo {volume} {66}},\ \bibinfo
  {pages} {015207} (\bibinfo {year} {2002})}\BibitemShut {NoStop}%
\bibitem [{\citenamefont {Beri}\ \emph {et~al.}(2005)\citenamefont {Beri},
  \citenamefont {Mannella}, \citenamefont {Luchinsky}, \citenamefont
  {Silchenko},\ and\ \citenamefont {McClintock}}]{Beri2005}%
  \BibitemOpen
  \bibfield  {author} {\bibinfo {author} {\bibfnamefont {S.}~\bibnamefont
  {Beri}}, \bibinfo {author} {\bibfnamefont {R.}~\bibnamefont {Mannella}},
  \bibinfo {author} {\bibfnamefont {D.~G.}\ \bibnamefont {Luchinsky}}, \bibinfo
  {author} {\bibfnamefont {A.~N.}\ \bibnamefont {Silchenko}}, \ and\ \bibinfo
  {author} {\bibfnamefont {P.~V.~E.}\ \bibnamefont {McClintock}},\ }\href
  {\doibase 10.1103/PhysRevE.72.036131} {\bibfield  {journal} {\bibinfo
  {journal} {Phys. Rev. E}\ }\textbf {\bibinfo {volume} {72}},\ \bibinfo
  {pages} {036131} (\bibinfo {year} {2005})}\BibitemShut {NoStop}%
\bibitem [{\citenamefont {{B{\'o}dai}}(2018)}]{Bodai2018}%
  \BibitemOpen
  \bibfield  {author} {\bibinfo {author} {\bibfnamefont {T.}~\bibnamefont
  {{B{\'o}dai}}},\ }\href@noop {} {\bibfield  {journal} {\bibinfo  {journal}
  {ArXiv e-prints}\ ,\ \bibinfo {pages} {arXiv:1808.06903}} (\bibinfo {year}
  {2018})}\BibitemShut {NoStop}%
\bibitem [{\citenamefont {Lindley}\ and\ \citenamefont
  {Schwartz}(2013)}]{Lindley2013}%
  \BibitemOpen
  \bibfield  {author} {\bibinfo {author} {\bibfnamefont {B.~S.}\ \bibnamefont
  {Lindley}}\ and\ \bibinfo {author} {\bibfnamefont {I.~B.}\ \bibnamefont
  {Schwartz}},\ }\href {\doibase https://doi.org/10.1016/j.physd.2013.04.001}
  {\bibfield  {journal} {\bibinfo  {journal} {Physica D}\ }\textbf {\bibinfo
  {volume} {255}},\ \bibinfo {pages} {22 } (\bibinfo {year}
  {2013})}\BibitemShut {NoStop}%
\bibitem [{\citenamefont {Tang}\ \emph {et~al.}(2017)\citenamefont {Tang},
  \citenamefont {Yuan}, \citenamefont {Wang}, \citenamefont {Zhu},\ and\
  \citenamefont {Ao}}]{Tang2017}%
  \BibitemOpen
  \bibfield  {author} {\bibinfo {author} {\bibfnamefont {Y.}~\bibnamefont
  {Tang}}, \bibinfo {author} {\bibfnamefont {R.}~\bibnamefont {Yuan}}, \bibinfo
  {author} {\bibfnamefont {G.}~\bibnamefont {Wang}}, \bibinfo {author}
  {\bibfnamefont {X.}~\bibnamefont {Zhu}}, \ and\ \bibinfo {author}
  {\bibfnamefont {P.}~\bibnamefont {Ao}},\ }\href {\doibase
  10.1038/s41598-017-15889-2} {\bibfield  {journal} {\bibinfo  {journal}
  {Scientific Reports}\ }\textbf {\bibinfo {volume} {7}},\ \bibinfo {pages}
  {15762} (\bibinfo {year} {2017})}\BibitemShut {NoStop}%
\bibitem [{\citenamefont {Zinn-Justin}(1996)}]{ZinnJustin1996}%
  \BibitemOpen
  \bibfield  {author} {\bibinfo {author} {\bibfnamefont {J.}~\bibnamefont
  {Zinn-Justin}},\ }\href@noop {} {\emph {\bibinfo {title} {Quantum Field
  Theory and Critical Phenomena}}}\ (\bibinfo  {publisher} {Oxford University
  Press},\ \bibinfo {address} {Oxford},\ \bibinfo {year} {1996})\BibitemShut
  {NoStop}%
\bibitem [{\citenamefont {Medeiros}\ \emph {et~al.}(2017)\citenamefont
  {Medeiros}, \citenamefont {Caldas}, \citenamefont {Baptista},\ and\
  \citenamefont {Feudel}}]{Medeiros2017}%
  \BibitemOpen
  \bibfield  {author} {\bibinfo {author} {\bibfnamefont {E.~S.}\ \bibnamefont
  {Medeiros}}, \bibinfo {author} {\bibfnamefont {I.}~\bibnamefont {Caldas}},
  \bibinfo {author} {\bibfnamefont {M.~S.}\ \bibnamefont {Baptista}}, \ and\
  \bibinfo {author} {\bibfnamefont {U.}~\bibnamefont {Feudel}},\ }\href
  {https://doi.org/10.1038/srep42351} {\bibfield  {journal} {\bibinfo
  {journal} {Scientific Reports}\ }\textbf {\bibinfo {volume} {7}},\ \bibinfo
  {pages} {42351 EP } (\bibinfo {year} {2017})}\BibitemShut {NoStop}%
\bibitem [{\citenamefont {Ashwin}\ \emph {et~al.}(2012)\citenamefont {Ashwin},
  \citenamefont {Wieczorek}, \citenamefont {Vitolo},\ and\ \citenamefont
  {Cox}}]{Ashwin2012}%
  \BibitemOpen
  \bibfield  {author} {\bibinfo {author} {\bibfnamefont {P.}~\bibnamefont
  {Ashwin}}, \bibinfo {author} {\bibfnamefont {S.}~\bibnamefont {Wieczorek}},
  \bibinfo {author} {\bibfnamefont {R.}~\bibnamefont {Vitolo}}, \ and\ \bibinfo
  {author} {\bibfnamefont {P.}~\bibnamefont {Cox}},\ }\href {\doibase
  10.1098/rsta.2011.0306} {\bibfield  {journal} {\bibinfo  {journal}
  {Philosophical Transactions of the Royal Society of London A: Mathematical,
  Physical and Engineering Sciences}\ }\textbf {\bibinfo {volume} {370}},\
  \bibinfo {pages} {1166} (\bibinfo {year} {2012})}\BibitemShut {NoStop}%
\bibitem [{\citenamefont {Steffen}\ and\ \citenamefont
  {et~al.}(2018)}]{Steffen2018}%
  \BibitemOpen
  \bibfield  {author} {\bibinfo {author} {\bibfnamefont {W.}~\bibnamefont
  {Steffen}}\ and\ \bibinfo {author} {\bibnamefont {et~al.}},\ }\href {\doibase
  10.1073/pnas.1810141115} {\bibfield  {journal} {\bibinfo  {journal}
  {Proceedings of the National Academy of Sciences}\ }\textbf {\bibinfo
  {volume} {115}},\ \bibinfo {pages} {8252} (\bibinfo {year}
  {2018})}\BibitemShut {NoStop}%
\bibitem [{\citenamefont {Gomez-Leal}\ \emph {et~al.}(2018)\citenamefont
  {Gomez-Leal}, \citenamefont {Kaltenegger}, \citenamefont {Lucarini},\ and\
  \citenamefont {Lunkeit}}]{Gomez2018}%
  \BibitemOpen
  \bibfield  {author} {\bibinfo {author} {\bibfnamefont {I.}~\bibnamefont
  {Gomez-Leal}}, \bibinfo {author} {\bibfnamefont {L.}~\bibnamefont
  {Kaltenegger}}, \bibinfo {author} {\bibfnamefont {V.}~\bibnamefont
  {Lucarini}}, \ and\ \bibinfo {author} {\bibfnamefont {F.}~\bibnamefont
  {Lunkeit}},\ }\href {\doibase https://doi.org/10.1016/j.icarus.2018.11.019}
  {\bibfield  {journal} {\bibinfo  {journal} {Icarus}\ } (\bibinfo {year}
  {2018}),\ https://doi.org/10.1016/j.icarus.2018.11.019}\BibitemShut {NoStop}%
\end{thebibliography}

%

\end{document}